\documentclass[10pt,a4paper,onecolumn]{article}
\usepackage{marginnote}
\usepackage{graphicx}
\usepackage{xcolor}
\usepackage{authblk,etoolbox}
\usepackage{titlesec}
\usepackage{calc}
\usepackage{tikz}
\usepackage{hyperref}
\hypersetup{colorlinks,breaklinks=true,
            urlcolor=[rgb]{0.0, 0.5, 1.0},
            linkcolor=[rgb]{0.0, 0.5, 1.0}}
\usepackage{caption}
\usepackage{tcolorbox}
\usepackage{amssymb,amsmath}
\usepackage{ifxetex,ifluatex}
\usepackage{seqsplit}
\usepackage{xstring}

\usepackage{float}
\let\origfigure\figure
\let\endorigfigure\endfigure
\renewenvironment{figure}[1][2] {
    \expandafter\origfigure\expandafter[H]
} {
    \endorigfigure
}

\usepackage{fixltx2e} 
\usepackage[
  backend=biber,
]{biblatex}
\bibliography{paper.bib}

\let\textttOrig=\texttt
\def\texttt#1{\expandafter\textttOrig{\seqsplit{#1}}}
\renewcommand{\seqinsert}{\ifmmode
  \allowbreak
  \else\penalty6000\hspace{0pt plus 0.02em}\fi}

\makeatletter
\let\href@Orig=\href
\def\href@Urllike#1#2{\href@Orig{#1}{\begingroup
    \def\Url@String{#2}\Url@FormatString
    \endgroup}}
\def\href@Notdoi#1#2{\def\tempa{#1}\def\tempb{#2}%
  \ifx\tempa\tempb\relax\href@Urllike{#1}{#2}\else
  \href@Orig{#1}{#2}\fi}
\def\href#1#2{%
  \IfBeginWith{#1}{https://doi.org}%
  {\href@Urllike{#1}{#2}}{\href@Notdoi{#1}{#2}}}
\makeatother

\usepackage[top=3.5cm, bottom=3cm, right=1.5cm, left=1.0cm,
            headheight=2.2cm, reversemp, includemp, marginparwidth=4.5cm]{geometry}

\titleformat{\section}
  {\normalfont\sffamily\Large\bfseries}
  {}{0pt}{}
\titleformat{\subsection}
  {\normalfont\sffamily\large\bfseries}
  {}{0pt}{}
\titleformat{\subsubsection}
  {\normalfont\sffamily\bfseries}
  {}{0pt}{}
\titleformat*{\paragraph}
  {\sffamily\normalsize}

\usepackage{fancyhdr}
\makeatletter
\let\ps@plain\ps@fancy
\fancyheadoffset[L]{4.5cm}
\fancyfootoffset[L]{4.5cm}

\definecolor{linky}{rgb}{0.0, 0.5, 1.0}

\newtcolorbox{repobox}
   {colback=red, colframe=red!75!black,
     boxrule=0.5pt, arc=2pt, left=6pt, right=6pt, top=3pt, bottom=3pt}

\newcommand{\ExternalLink}{%
   \tikz[x=1.2ex, y=1.2ex, baseline=-0.05ex]{%
       \begin{scope}[x=1ex, y=1ex]
           \clip (-0.1,-0.1)
               --++ (-0, 1.2)
               --++ (0.6, 0)
               --++ (0, -0.6)
               --++ (0.6, 0)
               --++ (0, -1);
           \path[draw,
               line width = 0.5,
               rounded corners=0.5]
               (0,0) rectangle (1,1);
       \end{scope}
       \path[draw, line width = 0.5] (0.5, 0.5)
           -- (1, 1);
       \path[draw, line width = 0.5] (0.6, 1)
           -- (1, 1) -- (1, 0.6);
       }
   }

\patchcmd{\@maketitle}{center}{flushleft}{}{}
\patchcmd{\@maketitle}{center}{flushleft}{}{}
\patchcmd{\@maketitle}{\LARGE}{\LARGE\sffamily}{}{}
\def\maketitle{{%
  
  \AB@maketitle}}
\makeatletter
\renewcommand\AB@affilsepx{ \protect\Affilfont}
\renewcommand\AB@affilnote[1]{{\bfseries #1}\hspace{3pt}}
\renewcommand{\affil}[2][]%
   {\newaffiltrue\let\AB@blk@and\AB@pand
      \if\relax#1\relax\def\AB@note{\AB@thenote}\else\def\AB@note{#1}%
        \setcounter{Maxaffil}{0}\fi
        \begingroup
        \let\href=\href@Orig
        \let\texttt=\textttOrig
        \let\protect\@unexpandable@protect
        \def\thanks{\protect\thanks}\def\footnote{\protect\footnote}%
        \@temptokena=\expandafter{\AB@authors}%
        {\def\\{\protect\\\protect\Affilfont}\xdef\AB@temp{#2}}%
         \xdef\AB@authors{\the\@temptokena\AB@las\AB@au@str
         \protect\\[\affilsep]\protect\Affilfont\AB@temp}%
         \gdef\AB@las{}\gdef\AB@au@str{}%
        {\def\\{, \ignorespaces}\xdef\AB@temp{#2}}%
        \@temptokena=\expandafter{\AB@affillist}%
        \xdef\AB@affillist{\the\@temptokena \AB@affilsep
          \AB@affilnote{\AB@note}\protect\Affilfont\AB@temp}%
      \endgroup
       \let\AB@affilsep\AB@affilsepx
}
\makeatother

\renewcommand\Affilfont{\sffamily\small\mdseries}
\ifnum 0\ifxetex 1\fi\ifluatex 1\fi=0 
  \usepackage[T1]{fontenc}
  \usepackage[utf8]{inputenc}

\else 
  \ifxetex
    \usepackage{mathspec}
    \usepackage{fontspec}

  \else
    \usepackage{fontspec}
  \fi
  \defaultfontfeatures{Ligatures=TeX,Scale=MatchLowercase}

\fi
\IfFileExists{upquote.sty}{\usepackage{upquote}}{}
\IfFileExists{microtype.sty}{%
\usepackage{microtype}
\UseMicrotypeSet[protrusion]{basicmath} 
}{}

\usepackage{hyperref}
\hypersetup{unicode=true,
            pdftitle={PyFstat: a Python package for continuous gravitational-wave data analysis},
            pdfborder={0 0 0},
            breaklinks=true}
\let\addcontentslineOrig=\addcontentsline
\def\addcontentsline#1#2#3{\bgroup
  \let\texttt=\textttOrig\addcontentslineOrig{#1}{#2}{#3}\egroup}
\let\markbothOrig\markboth
\def\markboth#1#2{\bgroup
  \let\texttt=\textttOrig\markbothOrig{#1}{#2}\egroup}
\let\markrightOrig\markright
\def\markright#1{\bgroup
  \let\texttt=\textttOrig\markrightOrig{#1}\egroup}

\usepackage{graphicx,grffile}
\makeatletter
\def\maxwidth{\ifdim\Gin@nat@width>\linewidth\linewidth\else\Gin@nat@width\fi}
\def\maxheight{\ifdim\Gin@nat@height>\textheight\textheight\else\Gin@nat@height\fi}
\makeatother
\setkeys{Gin}{width=\maxwidth,height=\maxheight,keepaspectratio}
\IfFileExists{parskip.sty}{%
\usepackage{parskip}
}{
\setlength{\parindent}{0pt}
\setlength{\parskip}{6pt plus 2pt minus 1pt}
}
\providecommand{\tightlist}{%
  \setlength{\itemsep}{0pt}\setlength{\parskip}{0pt}}
\ifx\paragraph\undefined\else
\let\oldparagraph\paragraph
\renewcommand{\paragraph}[1]{\oldparagraph{#1}\mbox{}}
\fi
\ifx\subparagraph\undefined\else
\let\oldsubparagraph\subparagraph
\renewcommand{\subparagraph}[1]{\oldsubparagraph{#1}\mbox{}}
\fi

\title{PyFstat: a Python package for continuous gravitational-wave data
analysis}

        \author[1]{David Keitel}
          \author[1]{Rodrigo Tenorio}
          \author[2]{Gregory Ashton}
          \author[3, 4]{Reinhard Prix}
    
      \affil[1]{Departament de Física, Institut d'Aplicacions Computacionals i de Codi
Comunitari (IAC3), Universitat de les Illes Balears, and Institut
d'Estudis Espacials de Catalunya (IEEC), Crta. Valldemossa km 7.5,
E-07122 Palma, Spain}
      \affil[2]{OzGrav, School of Physics \& Astronomy, Monash University, Clayton 3800,
Victoria, Australia}
      \affil[3]{Max-Planck-Institut für Gravitationsphysik (Albert-Einstein-Institut),
D-30167 Hannover, Germany}
      \affil[4]{Leibniz Universität Hannover, D-30167 Hannover, Germany}
  \date{\vspace{-7ex}}

\begin{document}
\maketitle

\marginpar{

  \begin{flushleft}
  \sffamily\small

  {\bfseries DOI:} \href{https://doi.org/10.21105/joss.03000}{\color{linky}{10.21105/joss.03000}}

  \vspace{2mm}

  {\bfseries Software}
  \begin{itemize}
    \setlength\itemsep{0em}
    \item \href{https://github.com/openjournals/joss-reviews/issues/3000}{\color{linky}{Review}} \ExternalLink
    \item \href{https://github.com/pyfstat/pyfstat}{\color{linky}{Repository}} \ExternalLink
    \item \href{https://doi.org/10.5281/zenodo.4660591}{\color{linky}{Archive}} \ExternalLink
  \end{itemize}

  \vspace{2mm}

  \par\noindent\hrulefill\par

  \vspace{2mm}

  {\bfseries Editor:} \href{https://danielskatz.org/}{Daniel S. Katz} \ExternalLink \\
  \vspace{1mm}
    {\bfseries Reviewers:}
  \begin{itemize}
  \setlength\itemsep{0em}
    \item \href{https://github.com/RobertRosca}{@RobertRosca}
    \item \href{https://github.com/khanx169}{@khanx169}
    \end{itemize}
    \vspace{2mm}

  {\bfseries Submitted:} 26 January 2021\\
  {\bfseries Published:} 06 April 2021

  \vspace{2mm}
  {\bfseries License}\\
  Authors of papers retain copyright and release the work under a Creative Commons Attribution 4.0 International License (\href{http://creativecommons.org/licenses/by/4.0/}{\color{linky}{CC BY 4.0}}).

  \end{flushleft}
}

\hypertarget{summary}{%
\section{Summary}\label{summary}}

Gravitational waves in the sensitivity band of ground-based detectors
can be emitted by a number of astrophysical sources, including not only
binary coalescences, but also individual spinning neutron stars. The
most promising signals from such sources, although as of 2020 not yet
detected, are the long-lasting, quasi-monochromatic `Continuous Waves'
(CWs). Many search methods have been developed and applied on LIGO (Aasi
et al. 2015) and Virgo (Acernese et al. 2015) data. See Prix (2009),
Riles (2017), and Sieniawska and Bejger (2019) for reviews of the field.

The \texttt{PyFstat} package provides tools to perform a range of CW
data analysis tasks. It revolves around the \(\mathcal{F}\)-statistic,
first introduced by Jaranowski, Krolak, and Schutz (1998): a
matched-filter detection statistic for CW signals described by a set of
frequency evolution parameters and maximized over amplitude parameters.
This has been one of the standard methods for LIGO-Virgo CW searches for
two decades. \texttt{PyFstat} is built on top of established routines in
\texttt{LALSuite} (LIGO Scientific Collaboration 2018) but through its
more modern \texttt{Python} interface it enables a flexible approach to
designing new search strategies.

Classes for various search strategies and target signals are contained
in three main submodules:

\begin{itemize}
\tightlist
\item
  \texttt{core}: The basic wrappers to \texttt{LALSuite}'s
  \(\mathcal{F}\)-statistic algorithm. End-users should rarely need to
  access these directly.
\item
  \texttt{grid\_based\_searches}: Classes to search over regular
  parameter-space grids.
\item
  \texttt{mcmc\_based\_searches}: Classes to cover promising
  parameter-space regions through stochastic template placement with the
  Markov Chain Monte Carlo (MCMC) sampler \texttt{ptemcee} (Vousden,
  Farr, and Mandel 2015).
\end{itemize}

Besides standard CWs from isolated neutron stars, \texttt{PyFstat} can
also be used to search for CWs from sources in binary systems (including
the additional orbital parameters), for CWs with a discontinuity at a
pulsar glitch, and for CW-like long-duration transient signals, e.g.,
from \emph{after} a pulsar glitch. Specialized versions of both
grid-based and MCMC-based search classes are provided for these
scenarios. Both fully-coherent and semi-coherent searches (where the
data is split into several segments for efficiency) are covered, and an
extension to the \(\mathcal{F}\)-statistic that is more robust against
single-detector noise artifacts (Keitel et al. 2014) is also supported.
While \texttt{PyFstat}'s grid-based searches do not compete with the
sophisticated grid setups and semi-coherent algorithms implemented in
various \texttt{LALSuite} programs, its main scientific use cases so far
are for the MCMC exploration of interesting parameter-space regions and
for the long-duration transient case.

\texttt{PyFstat} was first introduced in Ashton and Prix (2018), which
remains the main reference for the MCMC-based analysis implemented in
the package. The extension to transient signals, which uses
\texttt{PyCUDA} (Klöckner et al. 2012) for speedup, is discussed in
detail in Keitel and Ashton (2018), and the glitch-robust search
approaches in Ashton, Prix, and Jones (2018).

Additional helper classes, utility functions, and internals are included
for handling the common Short Fourier Transform (SFT) data format for
LIGO data, simulating artificial data with noise and signals in them,
and plotting results and diagnostics. Most of the underlying
\texttt{LALSuite} functionality is accessed through SWIG wrappings
(Wette 2020) though for some parts, such as the SFT handling, we still
(as of the writing of this paper) call stand-alone \texttt{lalapps}
executables. Completing the backend migration to pure SWIG usage is
planned for the future.

The source of \texttt{PyFstat} is hosted on
\href{https://github.com/PyFstat/PyFstat/}{GitHub}. The repository also
contains an automated test suite and a set of introductory example
scripts. Issues with the software can be submitted through GitHub and
pull requests are always welcome. \texttt{PyFstat} can be installed
through pip, conda or docker containers. Documentation in html and pdf
formats is available from
\href{https://readthedocs.org/projects/pyfstat/}{readthedocs.org} and
installation instructions can be found there or in the
\href{https://github.com/PyFstat/PyFstat/blob/master/README.md}{README}
file. PyFstat is also listed in the Astrophysics Source Code Library as
\href{https://ascl.net/2102.027}{ascl:2102.027}.

\hypertarget{statement-of-need}{%
\section{Statement of need}\label{statement-of-need}}

The sensitivity of searches for CWs and long-duration transient GWs is
generally limited by computational resources, as the required number of
matched-filter templates increases steeply for long observation times
and wide parameter spaces. The C-based \texttt{LALSuite} library (LIGO
Scientific Collaboration 2018) contains many sophisticated search
methods with a long development history and high level of optimization,
but is not very accessible for researchers new to the field or for
students; nor is it convenient for rapid development and integration
with modern technologies like GPUs or machine learning. Hence,
\texttt{PyFstat} serves a dual function of (i) making \texttt{LALSuite}
CW functionality more easily accessible through a \texttt{Python}
interface, thus facilitating the new user experience and, for
developers, the exploratory implementation of novel methods; and (ii)
providing a set of production-ready search classes for use cases not yet
covered by \texttt{LALSuite} itself, most notably for MCMC-based
followup of promising candidates from wide-parameter-space searches.

So far, \texttt{PyFstat} has been used for

\begin{itemize}
\tightlist
\item
  the original proposal of MCMC followup for CW candidates (Ashton and
  Prix 2018);
\item
  developing glitch-robust CW search methods (Ashton, Prix, and Jones
  2018);
\item
  speeding up long-transient searches with GPUs (Keitel and Ashton
  2018);
\item
  followup of candidates from all-sky searches for CWs from sources in
  binary systems, see Covas and Sintes (2020) and Abbott et al. (2021);
\item
  studying the impact of neutron star proper motions on CW searches
  (Covas 2020).
\end{itemize}

\hypertarget{acknowledgements}{%
\section{Acknowledgements}\label{acknowledgements}}

We acknowledge contributions to the package from Karl Wette, Sylvia Zhu
and Dan Foreman-Mackey; as well as helpful suggestions by John T.
Whelan, Luca Rei, and the LIGO-Virgo-KAGRA Continuous Wave working
group. D.K. and R.T. are supported by European Union FEDER funds; the
Spanish Ministerio de Ciencia, Innovación y Universidades and Agencia
Estatal de Investigación grants
PID2019-106416GB-I00/AEI/10.13039/501100011033, RED2018-102661-T,
RED2018-102573-E, FPA2017-90687-REDC, FPU 18/00694, and BEAGAL 18/00148
(cofinanced by the Universitat de les Illes Balears); the Comunitat
Autonoma de les Illes Balears through the Direcció General de Política
Universitaria i Recerca with funds from the Tourist Stay Tax Law ITS
2017-006 (PRD2018/24) and the Conselleria de Fons Europeus, Universitat
i Cultura; the Generalitat Valenciana (PROMETEO/2019/071); and EU COST
Actions CA18108, CA17137, CA16214, and CA16104. This paper has been
assigned document number LIGO-P2100008.

\hypertarget{references}{%
\section*{References}\label{references}}
\addcontentsline{toc}{section}{References}

\hypertarget{refs}{}
\leavevmode\hypertarget{ref-TheLIGOScientific:2014jea}{}%
Aasi, J., B. P. Abbott, R. Abbott, and others. 2015. ``Advanced LIGO.''
\emph{Class. Quant. Grav.} 32: 074001.
\url{https://doi.org/10.1088/0264-9381/32/7/074001}.

\leavevmode\hypertarget{ref-Abbott:2020mev}{}%
Abbott, R., T. D. Abbott, S. Abraham, and others. 2021. ``All-sky search
in early O3 LIGO data for continuous gravitational-wave signals from
unknown neutron stars in binary systems.'' \emph{Phys. Rev. D} 103 (6):
064017. \url{https://doi.org/10.1103/PhysRevD.103.064017}.

\leavevmode\hypertarget{ref-TheVirgo:2014hva}{}%
Acernese, F., M. Agathos, K. Agatsuma, and others. 2015. ``Advanced
Virgo: a second-generation interferometric gravitational wave
detector.'' \emph{Class. Quant. Grav.} 32 (2): 024001.
\url{https://doi.org/10.1088/0264-9381/32/2/024001}.

\leavevmode\hypertarget{ref-Ashton:2018ure}{}%
Ashton, Gregory, and Reinhard Prix. 2018. ``Hierarchical multistage MCMC
follow-up of continuous gravitational wave candidates.'' \emph{Phys.
Rev. D} 97 (10): 103020.
\url{https://doi.org/10.1103/PhysRevD.97.103020}.

\leavevmode\hypertarget{ref-Ashton:2018qth}{}%
Ashton, Gregory, Reinhard Prix, and D.I. Jones. 2018. ``A semicoherent
glitch-robust continuous-gravitational-wave search method.'' \emph{Phys.
Rev. D} 98 (6): 063011.
\url{https://doi.org/10.1103/PhysRevD.98.063011}.

\leavevmode\hypertarget{ref-Covas:2020hcy}{}%
Covas, P. B. 2020. ``Effects of proper motion of neutron stars on
continuous gravitational-wave searches.'' \emph{Mon. Not. Roy. Astron.
Soc.} 500 (4): 5167--76. \url{https://doi.org/10.1093/mnras/staa3624}.

\leavevmode\hypertarget{ref-Covas:2020nwy}{}%
Covas, P. B., and Alicia M. Sintes. 2020. ``First all-sky search for
continuous gravitational-wave signals from unknown neutron stars in
binary systems using Advanced LIGO data.'' \emph{Phys. Rev. Lett.} 124
(19): 191102. \url{https://doi.org/10.1103/PhysRevLett.124.191102}.

\leavevmode\hypertarget{ref-Jaranowski:1998qm}{}%
Jaranowski, Piotr, Andrzej Krolak, and Bernard F. Schutz. 1998. ``Data
analysis of gravitational - wave signals from spinning neutron stars. 1.
The Signal and its detection.'' \emph{Phys. Rev. D} 58: 063001.
\url{https://doi.org/10.1103/PhysRevD.58.063001}.

\leavevmode\hypertarget{ref-Keitel:2018pxz}{}%
Keitel, David, and Gregory Ashton. 2018. ``Faster search for long
gravitational-wave transients: GPU implementation of the transient
\(\mathcal F\)-statistic.'' \emph{Class. Quant. Grav.} 35 (20): 205003.
\url{https://doi.org/10.1088/1361-6382/aade34}.

\leavevmode\hypertarget{ref-Keitel:2013wga}{}%
Keitel, David, Reinhard Prix, Maria Alessandra Papa, Paola Leaci, and
Maham Siddiqi. 2014. ``Search for continuous gravitational waves:
Improving robustness versus instrumental artifacts.'' \emph{Phys. Rev.
D} 89 (6): 064023. \url{https://doi.org/10.1103/PhysRevD.89.064023}.

\leavevmode\hypertarget{ref-Kloeckner:2012pyc}{}%
Klöckner, Andreas, Nicolas Pinto, Yunsup Lee, B. Catanzaro, Paul Ivanov,
and Ahmed Fasih. 2012. ``PyCUDA and PyOpenCL: A Scripting-Based Approach
to GPU Run-Time Code Generation.'' \emph{Parallel Computing} 38 (3):
157--74. \url{https://doi.org/10.1016/j.parco.2011.09.001}.

\leavevmode\hypertarget{ref-lalsuite}{}%
LIGO Scientific Collaboration. 2018. ``LIGO Algorithm Library -
LALSuite.'' free software (GPL).
\url{https://doi.org/10.7935/GT1W-FZ16}.

\leavevmode\hypertarget{ref-Prix:2009oha}{}%
Prix, Reinhard. 2009. ``Gravitational Waves from Spinning Neutron
Stars.'' In \emph{Neutron Stars and Pulsars}, edited by Werner Becker,
357:651--85. Astrophys. Space Sci. Lib. Berlin Heidelberg: Springer.
\url{https://doi.org/10.1007/978-3-540-76965-1_24}.

\leavevmode\hypertarget{ref-Riles:2017evm}{}%
Riles, Keith. 2017. ``Recent searches for continuous gravitational
waves.'' \emph{Mod. Phys. Lett. A} 32 (39): 1730035.
\url{https://doi.org/10.1142/S021773231730035X}.

\leavevmode\hypertarget{ref-Sieniawska:2019hmd}{}%
Sieniawska, Magdalena, and Michał Bejger. 2019. ``Continuous
gravitational waves from neutron stars: current status and prospects.''
\emph{Universe} 5 (11): 217.
\url{https://doi.org/10.3390/universe5110217}.

\leavevmode\hypertarget{ref-Vousden:2015pte}{}%
Vousden, W. D., W. M. Farr, and I. Mandel. 2015. ``Dynamic temperature
selection for parallel tempering in Markov chain Monte Carlo
simulations.'' \emph{Mon. Not. Roy. Astron. Soc.} 455 (2): 1919--37.
\url{https://doi.org/10.1093/mnras/stv2422}.

\leavevmode\hypertarget{ref-Wette:2020air}{}%
Wette, Karl. 2020. ``SWIGLAL: Python and Octave interfaces to the
LALSuite gravitational-wave data analysis libraries.'' \emph{SoftwareX}
12: 100634. \url{https://doi.org/10.1016/j.softx.2020.100634}.

\end{document}